      [1999/12/01 v1.4c Il Nuovo Cimento]
\documentclass{cimento}
\usepackage{pstricks,pst-node,pst-text,pst-3d}
\usepackage{amssymb}
\usepackage{amsmath}
\newcommand{\bq}{\begin{eqnarray}}
\newcommand{\eq}{\end{eqnarray}}
\def\sumint{\int \! \!\ \! \! \! \! \!\ \! \! \!\! \!\sum}
\newcommand{\blf}[1]{\bf  {\tilde #1}}
\newcommand{\bm}[1] {\mbox{\boldmath{$#1$}}}
\def\psla{ \slash \! \! \!}
\def\Psla{ \slash \! \!\! \!}
\def\Tr{\rm Tr}
\def\be{\begin{eqnarray} &&}
\def\nonu{\nonumber \\ &&}
\def\ee{\end{eqnarray}}

\title{
{ Transversity studies
with a polarized $^3$He target }
 }
\author{A.~Del Dotto\from{roma1}\ETC, E.~Pace\from{roma2},
G.~Salm\`e\from{roma1} \atque
S.~Scopetta\from{pg}}
\instlist{
\inst{roma2} Phys. Dept. Univ. of Rome ''Tor Vergata'' and INFN Sezione 
di ''Tor Vergata'', Via della Ricerca Scientifica 1, I-00133 Rome, Italy.
\inst{roma1} INFN, Sezione di Roma, P.le A. Moro 2, 00185 Rome, Italy.
\inst{pg} Dipartimento di Fisica, Universit\`a degli Studi
di Perugia and INFN, Sezione di Perugia, via A. Pascoli,
I-06100 Perugia, Italy.
}
\PACSes{\PACSit{13.60.Hb}
{13.85.Hd} --
{13.85.Ni} }
\begin{document}

\maketitle

\begin{abstract} {A realistic study of the SiDIS process $\vec{^3He}(e,e'\pi)X$
in the Bjorken limit is briefly reviewed, showing that the nuclear effects,
present  in the extraction of the neutron information, are largely under
theoretical control, within an Impulse-Approximation approach.  In view of the
forthcoming experimental data, we shortly present a novel Poincar\'e covariant
description of the nuclear target, implementing  a Light-Front analysis at
finite $Q^2$, within the Bakamijan-Thomas construction of the Poincar\'e
generators. Furthermore, as a by-product of the   introduction of a  
Light-Front  spin-dependent spectral function {for a J=1/2 system}, we
straightforwardly extend our analysis  to the quark-quark correlator,
obtaining   three new exact relations between the six leading-twist
Transverse-Momentum  Dependent   distributions.} \end{abstract}

\section{Introduction}

Due to ongoing and forthcoming challenging measurements
in various Laboratories, 
the partonic structure of transversely polarized hadrons,
not accessible in inclusive experiments,
is 
widely studied 
by the hadronic Physics Community.
In particular,
semi-inclusive deep inelastic scattering (SiDIS)
is one of the 
processes proposed to access transversity observables.
The theoretical description of SiDIS 
implies a complicated formalism,
accounting for the transverse motion of the quarks
in the target \cite{bdr,sidis,D'Alesio:2007jt,Pasquini:2008ax}.
In particular, the non-perturbative effects of
the intrinsic transverse momentum $\vec{k}_{T}$ of the quarks inside 
the nucleon may induce significant hadron azimuthal asymmetries 
\cite{Mulders:1995dh, Cahn:1978se}. In order to describe such quantities, a
central role is played by
the Transverse-Momentum Dependent parton distributions (TMDs), that depend,
besides on the fraction $x$ 
of longitudinal momentum of the target carried by the parton, 
as the standard parton distributions (PDs),
 also on its intrinsic 
transverse momentum.
The number of TMDs, six at leading twist, is fixed by counting the 
scalar quantities allowed by hermiticity, parity and time-reversal invariance. 
However, the existence of 
leading twist
final state interactions (FSI) allows for 
two additional time-reversal odd functions~\cite{Brodsky:2002cx},
namely, the Sivers and the Boer-Mulders functions
\cite{sivers, Boer:1997nt}. 

The experimental scenario which arises from the analysis
of SiDIS off transversely polarized proton and deuteron targets is puzzling,
showing a strong flavor dependence \cite{hermes,compass}.
With the aim { of} extracting the neutron information and 
clarifying the situation,
a measurement of SiDIS
off transversely polarized $^3$He has been proposed \cite{bro},
and an experiment, planned to measure azimuthal asymmetries
in the production of leading $\pi^\pm$  from transversely
polarized $^3$He, has been just completed at Jefferson Lab (JLab), 
with a beam energy of 6 GeV
\cite{prljlab}. Another experiment
will be performed after  upgrading the JLab beam to 12 GeV \cite{jlab12,alessio}.
In this contribution,   a   brief review of 
the realistic analysis of SiDIS
off transversely polarized $^3$He, as presented in Ref.
\cite{mio}, is given and it is extended in order  to implement  both the 
 Poincar\'e covariant treatment of the nuclear target and 
 the finite values of the actual four-momentum transfer $Q^2$. In particular, 
the Light-front (LF) spin-dependent spectral function for a generic J=1/2 target
is introduced, exploiting
 the Bakamijan-Thomas (BT) construction of the Poincar\'e generators 
 (see, e.g.
\cite{KP} for a detailed review) and it is 
 applied to the quark-quark correlator, 
 obtaining new relations between the six leading-twist TMDs.

\section{A realistic  description of  $^3$He Single Spin Asymmetries}
\label{uno}

In Ref. \cite{mio}, within an  impulse approximation 
(IA) framework, 
the formal expressions of the
Collins \cite{coll} and Sivers \cite{sivers} contributions to the azimuthal
Single Spin Asymmetry (SSA) for the production
of leading pions off $^3$He have been derived. In particular, 
   the initial transverse momentum of the struck quark has been included, 
and,  
 in order to give  a realistic 
description of the nuclear dynamics, $^3$He SSAs have been calculated by using 
the nucleon spin-dependent spectral function (SF),
that yields the probability distribution to find a nucleon with
given  three-momentum and missing energy inside a target nucleus 
with polarization ${\bf S}_A$. The  adopted SF
was the one obtained in Ref. \cite{over} { within a non relativistic framework
and} corresponding to 
the AV18 nucleon-nucleon interaction \cite{av18}.  
Moreover,
the nucleonic structure, i.e. PDs and fragmentation functions, 
has been described
by proper parametrizations of data \cite{ans}
or suitable model calculations
\cite{model}.
In Ref. \cite{mio}, the crucial issue of extracting
the neutron information from $^3 \vec{\rm He}$ data
has been also discussed and it was showing that, even for SiDIS, where  
fragmentation functions are present beside PD's, 
one can adopt the same 
extraction scheme already successfully 
applied to DIS. In this case, in order to take
 effectively into account   
the momentum and energy distributions
of the polarized bound nucleons in $^3$He,
it was adopted a model independent procedure, based
on the realistic evaluation
of the proton and neutron polarizations in $^3$He
\cite{old},   $p_p$ and $p_n$  respectively.
According to this extraction procedure, the neutron SSA,
$A^i_n$, is obtained from the nuclear one, $A^{exp,i}_3$,
as follows
\begin{equation}
A^i_n \simeq {1 \over p_n d_n} \left ( A^{exp,i}_3 - 2 p_p d_p
A^{exp,i}_p \right )~,
\label{extr}
\end{equation}
where $d_n(d_p)$ is the neutron (proton) dilution
factor. For a neutron $n$
(proton $p$) in $^3$He, the diluition factor, experimentally known, is given by
\begin{eqnarray}
d_{n(p)}(x,z)=
{\sum_q e_q^2
f^{q,{n(p)}} 
\left ( x \right )
D^{q,h} \left ( z  \right )
\over
\sum_{N=p,n}
\sum_q e_q^2
f^{q,N} 
( x )
D^{q,h} 
\left ( z \right )
}~,
\label{dilut}
\end{eqnarray}
where $ f^{q,N} ( x )$ are  the standard parton
distributions and $D^{q,h} 
\left ( z \right )$
the fragmentation functions with $z=E_h/\nu$
  (see \cite{mio} for details). 
In Eq. (\ref{extr}), the nucleon effective polarizations,
$p_p$ and $p_n$ take properly care of  all the effects in
$^3$He, like
Fermi motion and binding, that produce  a depolarization
of the bound neutron.
In Ref. \cite{mio}, it has been shown that,
for any experimentally relevant $x$ and $z$,
the extraction
scheme of
Eq. (\ref{extr}) is quite effective.
This important result is due to the peculiar kinematics
of the JLab experiments, which helps in two ways.
Firstly, it  favors the emission of fast  pions from the current-quark fragmentation, 
($z$ has been chosen in the range $0.45 < z < 0.6$,
which means that only high-energy pions are observed).
Secondly, the pions are detected in a narrow cone around the direction
of the momentum transfer, making  
 nuclear effects in the fragmentation 
functions  rather small \cite{mio}. Then, the leading nuclear effects result to
be the ones 
 affecting the parton distributions, as already found
in inclusive DIS \cite{old}, and they can be taken into account
in the usual way, i.e. by using Eq. (\ref{extr}) for the extraction of the
neutron information. 
In conclusion,  Eq. (\ref{extr}) represents a valuable tool
for the experimental data analysis  \cite{prljlab,jlab12}.

While the analysis in \cite{mio} was performed
assuming the experimental set-up of the experiment
described in Ref. \cite{prljlab}, but using DIS kinematics, an
analysis at the actual, finite values of the momentum
and energy transfers, $Q^2$ and $\nu$,  is in progress 
\cite{new}, and some of the results are anticipated in the next Section.
Moreover, in order to improve the relativized framework where the non
relativistic SF of Ref. \cite{over} was embedded,   the new
analysis is based on a fully 
Poincar\'e covariant description of the nuclear
  effects,.by introducing a LF spin-dependent 
  SF (see below). 

\section{A Light-Front description of  $^3 \vec{\rm He}$}
\label{due}
In the previous Section, it has been
reported that the extraction procedure of the neutron
information, successfully applied in DIS,
nicely works  
also in SiDIS, 
though
the calculation has been performed \cite{mio} in the Bjorken limit,
using a non relativistic  spin-dependent SF.
Therefore, it is natural to test the extraction procedure
in the  actual JLab kinematics, rather far
from the asymptotic one, and to investigate  
relativistic effects at nuclear level, in a more consistent way.
In order to achieve this, we adopt  the
Relativistic Hamiltonian Dynamics (RHD) framework \cite{dirac,KP}, suitable for
describing an interacting system in a Poincar\'e covariant way.

The RHD,
introduced by Dirac \cite{dirac}, can be combined with the Bakamijan-Thomas 
explicit construction of the Poincar\'e generators \cite{bt} for 
obtaining a description of SiDIS off $^3$He
which :
{\it i)} is
fully Poincar\'e covariant;
{\it ii)} has a fixed number of on-mass-shell constituents;
{\it iii)} allows one to use the Clebsch-Gordan coefficients
 to decompose the wave function (a benefit from the BT construction).
Obviously, this approach
  is not explicitly covariant, since locality is lost.
Between the three possible forms of RHD, the 
{\it Light-Front}  one has
several advantages:
{\it i)} seven Kinematical generators:
three LF-boosts (at variance with the dynamical nature of
the Instant-form boosts),  three components of the LF-momentum, $\tilde {\bf
P}\equiv \{P^+, {\bf P}_\perp\}$, and the 
rotation around the z-axis;
{\it ii)} 
the LF-boosts have a subgroup structure, so that the
separation 
of the intrinsic motion { is trivially achieved}
(as in the NR case);
{\it iii)} 
$P^+\geq 0$ leads to a meaningful Fock expansion, in presence of 
massive boson exchanges;
{\it iv)} there are
no square roots in the operator $P^-$, propagating the
state in the LF-time;
{\it v)} 
the Infinite Momentum frame description of DIS is easily included.
The main drawback is that the
transverse LF-rotations are dynamical, but within the BT construction,
one can define a {\it kinematical}, intrinsic
angular momentum, reobtaining the rotational invariance even in presence of a
truncated Fock space. { This  is a peculiar feature of  RHD, at variance 
with Quantum Field Theory where infinite degrees of freedom are present due to
 the locality. } 
The LF Hamiltonian Dynamics (LFHD) framework has been proven
to be very suitable for phenomenological studies (see, e.g. \cite{deut,nucl}).

For SiDIS, the nuclear hadronic tensor plays a central
role, and in IA it reads
\be
{{ 
{\cal W}^{\mu\nu}(Q^2,x_B,z,\tau_{hf}, \hat{\bf h},S_{He})}}  \propto 
 \sum_{\sigma,\sigma'}\sum_{\tau_{hf}} 
 \left.\sumint \right._{\epsilon^{min}_S}^{\epsilon^{max}_S}{~d
\epsilon_{S} }
\int_{M^2_N}^{(M_X-M_S)^2} dM^2_f 
\int_{\xi_{lo}}^{\xi_{up}} {d\xi\over (2\pi)^3}
~\times \nonu ~~~~{1\over \xi^2 (1-\xi)}
\int_{P^{min}_\perp}^{P^{max}_\perp}{d
P_\perp\over sin\theta   }~ (P^+ +q^+ - h^+)
~
{{
w^{\mu\nu}_{\sigma\sigma'}\left(\tau_{hf},{\blf q},{\blf h},{\blf P}\right)
}}
{{
{\cal P}^{\tau_{hf}}_{\sigma'\sigma}({\bf k},\epsilon_S,S_{He})
}}~.
\label{cento}
\ee
All the formal steps for obtaining Eq. (\ref{cento}) and  the explicit expression of the
integration limits will be
reported elsewhere \cite{new}. 
Let us describe here only the two
crucial terms appearing in the above equation.
The first one  is 
$ w^{\mu\nu}_{\sigma\sigma'}
\left(\tau_{hf},{\blf q},{\blf h},{\blf P}\right)$,
the SiDIS nucleonic tensor, that depends
not only on its spins $\sigma,\sigma'$ and  its LF-momentum,
$
{\blf P}
$,
but also on the isospin 
$\tau_{hf}$
and LF-momentum
${\blf h}$ of the produced pseudo-scalar meson.
The second one  is
$
{\cal P}^{\tau_{hf}}_{\sigma'\sigma}({\bf k},\epsilon_S,S_{He})~,
$
i.e., the LF spin-dependent SF, describing a nucleon with 
Cartesian momentum ${\bf k}$  inside 
a $^3$He with polarization
$S_{He}$,
when the spectator pair has an excitation energy $\epsilon_S$.
The LF spectral function is
defined  as
\be
  {\cal P}^{\tau}_{\sigma'\sigma}({\blf k},\epsilon_{S},S_{He})
\propto
~\sum_{\sigma_1 \sigma'_1} 
D^{{1 \over 2}} [{\cal R}_M^\dagger ({\blf
k})]_{\sigma'\sigma'_1}~
{{
{\cal S}^{\tau}_{\sigma'_1\sigma_1}({\blf k},\epsilon_{S},S_{He})
}}
D^{{1 \over 2}} [{\cal R}_M ({\blf
k})]_{\sigma_1\sigma}~,
\label{LFSF}\ee
with $
D^{{1 \over 2}} [{\cal R}_M ({\blf k})]=~
 {m+k^+-\imath 
{\bm \sigma} \cdot (\hat z \times
{\bf k}_{\perp}) /\sqrt{\left ( m +k^+ \right )^2 +|{\bf
k}_{\perp}|^2}}
$ the unitary Melosh Rotations 
and the instant-form SF given by
\be
{\cal S}^{\tau}_{\sigma'_1\sigma_1}({\blf k},\epsilon_{S},S_{He})
=
\sum_{J_S J_{zS}\alpha}\sum_{T_{S}\tau_{S} } ~\langle   T_{S},
\tau_{S},\alpha,\epsilon_{S} J_S J_{zS};\sigma'_1;\tau, {\bf k}|
\Psi_{0} S_{He} \rangle 
~\times \nonu 
\langle S_{He},
\Psi_0|{\bf k}\sigma_1\tau_; J_S J_{zS} 
\epsilon_{S}, \alpha, T_{S}, \tau_{S}\rangle
=~\left[ 
{{
B_{0,S_{He}}^{\tau}(|{\bf k}|,E)
}}
~+~
{\bm \sigma} \cdot {\bf f}^{\tau}_{S_{He}}
({\bf k},E) \right]_{\sigma'_1\sigma_1}~,
\label{ISF}\ee
where
$
{\bf f}^{\tau}_{S_{He}}({\bf k},E) ~=~
{\bf S}_A~
{{
B_{1,S_{He}}^{\tau}(|{\bf k}|,E)}}~+~\hat{k}~(\hat{k}
\cdot {\bf S}_A)~ 
{{
B_{2,S_{He}}^{\tau}(|{\bf k}|,E)
}}~.
$
It is tempting to approximate the instant-form SF by the one
evaluated in a non relativistic framework, since the BT construction {imposes
the same constraints adopted for the non relativistic
 Hamiltonian} \cite{KP}.
 
From Eqs. (\ref{LFSF}) and (\ref{ISF}), it follows that, remarkably, 
 the constituent SF for a J=1/2 
system
is given in terms of only {\it three} independent functions 
(${{B_i}}$, with ${i=0,1,2}$), within a RHD framework.

From the phenomenological side, it will be very important
to use the LF nuclear tensor, Eq. (\ref{cento}),
to evaluate the SSAs and to figure out whether or not
the proposed extraction procedure 
still holds
in the LF analysis. 
{The preliminary calculations clearly indicate}
that the effect of
considering the 
integration limits
evaluated in the actual JLab kinematics at 6 GeV,
and not in the Bjorken limit, as previously done,
is negligible.
From this point of view,
the situation will become even better
in the experiments planned at 12 GeV.
\section{The Light-Front quark-quark correlator}
\label{tre}
In general, the six leading-twist TMDs for a $J =1/2$ system with 4-momentum $P$
and polarization $S$ 
are introduced as a proper parametrization  of 
the so called quark-quark correlator, $\Phi(k, P, S)$ 
for a quark of 4-momentum $k$, in order to fulfill all the general symmetries
(see, e.g., Ref. \cite{bdr}). In particular, {one can starts 
with the following combinations of Dirac structures and scalar functions, $A_i$
and $\tilde A_i$,}  
\be
  \Phi(k, P, S) =\frac12
  \left\{ 
    { {A_1}} 
\, \Psla{P} +
    { {A_{2}}} 
\, S_L \, \gamma_5 \, \Psla{P} +
    { {A_3}} 
\, \Psla{P} \, \gamma_5 \, \psla{S}_\perp
  \right.
  \nonu +
  \left.
    \frac1{M} \, 
{{\widetilde{A}_1}} 
\, \vec{k}_\perp{\cdot}\vec{S}_\perp \,
    \gamma_5 \Psla{P}
    +
{{\widetilde{A}_1}} 
\, \frac{S_L}{M} \,
    \Psla{P} \, \gamma_5 \, \psla{k}_\perp
  +
    \frac1{M^2} \, 
{{\widetilde{A}_1}} 
\, \vec{k}_\perp{\cdot}\vec{S}_\perp \,
    \Psla{P} \, \gamma_5 \, \psla{k}_\perp
  \right\}~.
\end{eqnarray}
Then, the  six leading-twist TMDs, {\bf $ f_1,~  g_{1L},~ g_{1T},~h_1,~h_{1L},~h_{1T}$,} are identified 
as follows:
\begin{subequations}
\be
  \frac1{2P^+} \, \Tr(\gamma^+ \Phi)
  =
{{  f_1}} 
\, ,
\\ &&
  \frac1{2P^+} \, \Tr(\gamma^+ \gamma_5 \Phi) =
  S_L \, 
{{  g_{1L}}} 
+
  \frac1{M} \, \vec{k}_\perp{\cdot}\vec{S}_\perp \, 
{{g_{1T} }}
\, ,
\\&&
  \frac1{2P^+} \, \Tr(i \sigma^{i+} \gamma_5 \Phi)=
  S_\perp^i \, 
{{h_1}} 
+
  \frac{S_L}{M}\, k_\perp^i \, 
{{h_{1L}^\perp}} 
-
  \frac1{M^2} ~ ( k_\perp^i k_\perp^j + {1 \over 2}
k_\perp^2 g_\perp^{ij}){S}_{\perp,j} ~ 
{{h_{1T}^\perp}}  
\, .
\ee
\end{subequations}
The {LF} SF introduced in the previous Section for obtaining 
a Poincar\'e
covariant description of the nucleon inside a polarized $^3$He, can be formally
adopted for describing a quark inside a polarized nucleon. Namely,
 {\it ceteris paribus},  the LF  { nucleon spectral function,
 ${\cal P}_N$,} is the analogous of  
$\Phi(k, P, S)$, within a Poincar\'e covariant description of the nucleon.
 Then, one is able to perform the following identification 
\begin{subequations}
\bq
 {1 \over 2} {\Tr}( {\cal P}_N I) & = & {{c}} ~ {{B_0}}
  = 
  {{f_1^{LF} }}~,
\label{mille1}
\\
{1 \over 2}  
  {\Tr}(  {\cal P}_N \sigma_z ) & = &
~S_{ z}
{{\left [ a ~ ({{B_1}} + {{B_2}} \cos^2 \theta) 
+ b \cos \theta {| {\bf k}_\perp |^2\over
k} {{B_2}} 
\right ]}}
\nonumber
\\
& + & {\bf S}_\perp \cdot {\bf k}_\perp
{{\left [ a ~ {{B_2}} {\cos \theta \over k}
+ b ~({{B_1}} +{{B_2}}~\sin^2 \theta ) \right ]}}
\nonumber
\\
& = &
  S_L ~ {{g_{1L}^{LF}}} +
  \frac1{M} ~ 
{\bf S}_\perp \cdot {\bf k}_\perp~
{{g_{1T}^{LF}}}~, 
\\
{1 \over 2}
{\Tr}(  {\cal P}_N \sigma_y ) & = &
S_{y} \, {{\left[ \left( a + {d  \over 2}
~  |{\bf k}_\perp|^2 
\right)
~{{B_1 }}
+ {1 \over 2} \left( a - b {k_\perp^2 \cos \theta \over k } \right) 
~{{B_2 }}
\right]
}}
\nonumber
\\
& + &
S_{z} 
k_y
~{{\left[ a ~ 
{\cos \theta \over k}
{{B_2}} 
- b 
( {{B_1}}+
{{B_2}} 
\cos^2 \theta 
)\right]}}
\nonumber
\\
& + &
~ \left ( k_{x} k_{y} 
{S}_{x} ~ 
- {1 \over 2}
k_\perp^2 
{S}_{y}  
\right )~
{{
\left[
\left(  {a \over k^2} -b ~ 
{\cos \theta\over k} \right)
{{B_2}} 
- d ~ {{B_1}} \right]}}
\nonumber
\\
& = &
  S_{y} ~ {{h_1^{LF}}} +
  \frac{S_L}{M}~ k_y ~ {{h_{1L}^{\perp LF}}} +
  \frac1{M^2} ~ \left ( k_x k_y S_{x} - {1 \over 2}
k_\perp^2 S_{y} \right )~ 
{{h_{1T}^{\perp LF}}}~.
\label{mille}
\eq
\end{subequations}
Therefore, the {{{\it six} TMDs}} 
depend actually upon {{{\it three} independent functions, within a LFHD
framework}}.
The quantities
{{$a,b,c$ and $d$, appearing
in Eqs. 
(\ref{mille1})-(\ref{mille}), 
are {\it kinematical} factors, 
{\it predicted} by the LF procedure (details will be given
 elsewhere \cite{pss}}}).

\section{Conclusions}
A realistic study of {{ $\vec{^3He}(e,e'\pi)X$ }}
in the Bjorken limit has been summarized,
and the preliminary results of its generalization
within
a Light-Front analysis at finite $Q^2$ have been shortly
discussed. 
The spin-dependent LF spectral function { for a J=1/2 system
has been presented} and, using
the BT construction of the Poincar\'e generators, 
an intriguing simplification
in the theoretical description of SiDIS
is found. Three exact relations are established between the six T-even TMDs.
As a future step, it will be investigated if   similar
relations can be found   in other theoretical frameworks and  eventually 
in the analysis of the experimental results.

A detailed analysis of these new relations 
will be given elsewhere \cite{pss}.   

\acknowledgments
This work has been partially supported by the Italian MUR through the {\it 
PRIN 2008}.
One of us, S.S., thanks the organizers of the Conference
for the kind invitation.


\begin{thebibliography}{0}

\bibitem{bdr}
\BY{Barone V., Drago A. \atque Ratcliffe P.}
{\it Phys.\ Rept.}, {\bf 359} (2002) 1. 

\bibitem{sidis}

\BY{Bacchetta A., Diehl M., Goeke K., Metz A., Mulders P.J. 
\atque Schlegel M.} 
{\it J. High Energy Phys.}
JHEP, {\bf 0702} (2007) 093.

\bibitem{D'Alesio:2007jt}
\BY{D'Alesio U. \atque Murgia F.} 
{\it Prog.\ Part.\ Nucl.\ Phys.},  {\bf 61} (2008) 394.
  
\bibitem{Pasquini:2008ax}
\BY{Pasquini B., Cazzaniga S. \atque Boffi S.} 
{\it Phys.\ Rev.\ D}, {\bf 78} (2008) 034025.

\bibitem{Mulders:1995dh}
\BY{Mulders P.J. \atque Tangerman R.D.} 
{\it Nucl.\ Phys.\ B},  {\bf 461} (1996) 197
  [Erratum-ibid.\  {\bf 484} (1997) 538].
  
\bibitem{Cahn:1978se}
\BY{Cahn  R.~N.} 
{\it Phys.\ Lett.\ B}, {\bf 78} (1978) 269.

\bibitem{Brodsky:2002cx}
\BY{Brodsky S.J., Hwang D.S. \atque Schmidt I.}
{\it Phys.\ Lett.\ B}, {\bf 530} (2002) 99.
  
  \bibitem{sivers}
\BY{Sivers D.W.} 1
{\it  Phys.\ Rev.\ D}, {\bf 41} (1992) 83; 
{\it  Phys.\ Rev.\ D},  {\bf 43} (1991) 261. 
  
\bibitem{Boer:1997nt}
\BY{Boer D. \atque Mulders P.J.} 
{\it  Phys.\ Rev.\ D},  {\bf 57} (1998) 5780.

\bibitem{hermes} 
\BY{Airapetian A. {\sl et al.}} [HERMES Collaboration]
{\it Phys. Rev. Lett.}, {\bf 94} (2005) 012002.


\bibitem{compass}
\BY{Alexakhin V. Y. {\sl et al.}} [COMPASS Collaboration]
{\it Phys. Rev. Lett.}, {\bf 94} (2005) 202002.

\bibitem{bro} \BY{Brodsky S.J. \atque Gardner S.} 
{\it Phys. Lett. B}, {\bf 643} (2006) 22.

\bibitem{prljlab} 
\BY{Qian X.} {\it et al.} 
{\it Phys. Rev. Lett.}, {\bf 107} (2011) 072003.

\bibitem{jlab12} \BY{Gao H. {\it et al.}} 
{\it Eur. Phys. J. Plus}, {\bf 126} (2011) 2. 
\bibitem{alessio} \BY{Cates G. {\sl et  al}}, E12-09-018, JLAB approved experiment, 
\verb1www.hallaweb.jlab.org/collab/PAC/PAC38/E12-09-018-SIDIS.pdf1; 
\BY{Del Dotto A.} Master Thesis, Universit\`a di Roma
``La Sapienza'', (2011) (unpublished).

\bibitem{mio} \BY{Scopetta S.} 
{\it Phys. Rev. D}, {\bf 75} (2007) 054005.

\bibitem{KP} \BY{ Keister B.D. \atque   Polyzou W.N.} {\it Adv. Nucl. Phys. } 
{\bf 21},  (1991) 225.

\bibitem{coll}
\BY{Collins J.C.} {\it Nucl. Phys. B}, {\bf 396} (1993) 161. 

\bibitem{over} 
\BY{Kievsky A., Pace E., Salm\`e G., \atque Viviani M.} 
{\it Phys. Rev. C}, {\bf 56} (1997) 64.

\bibitem{av18} 
\BY{Wiringa R.B., Stocks V.G.J. \atque Schiavilla R.} 
{\it Phys. Rev. C}, {\bf 51} (1995) 38.
 
\bibitem{ans}
\BY{Anselmino M., Boglione M., D'Alesio U., Kotzinian A., 
Murgia F. \atque Prokudin A.} 
{\it Phys.\ Rev. D}, {\bf 71} (2005) 074006;
{\it Phys.\ Rev. D}, {\bf 72} (2005) 094007
[2005 Erratum-ibid.\  D {\bf 72} 099903].

\bibitem{model} 
\BY{Amrath D., Bacchetta A. \atque Metz A.} {\it Phys. Rev. D}, {\bf 71}
(2005) 114018. 

\bibitem{old} 
\BY{Ciofi degli Atti C., Scopetta S., Pace E. 
\atque Salm\`e G.}
{\it Phys. Rev. C} {\bf 48} (1993) 968. 

\bibitem{new} \BY{Del Dotto A., Salm\`e G. \atque Scopetta S.} in preparation.

\bibitem{dirac} \BY{Dirac P.A.M.} {\it Rev. Mod. Phys.}, {\bf 21} (1949) 392.

\bibitem{bt} \BY{Bakamijan B. \atque Thomas L.H.}  
{\it Phys. Rev.}, {\bf 92} (1953) 1300. 

\bibitem{deut} \BY{Lev F. M., Pace E. \atque Salm\`e G.} {\it Phys. Rev.} {\bf
C 62}, (2000) 064004.

\bibitem{nucl} \BY{Cardarelli F., Pace E., Salm\`e G. \atque Simula S.} 
{\it Phys. Lett.} {\bf B 357}, (1995) 267.

\bibitem{pss} \BY{Pace E., Salm\`e G. \atque Scopetta S.} 
in preparation.
\end{thebibliography}
\end{document}